\documentclass[epjST]{svjour}
\usepackage{graphicx,amssymb,amsmath,bbm,amsfonts}
\usepackage{amsmath}
\usepackage{amsfonts}
\usepackage{amssymb}
\usepackage[dvips,usenames]{color}
\usepackage{graphicx}
\newcommand{\ket}[1]{\vert #1 \rangle}

\newcommand{\ketbra}[2]{\vert #1 \rangle \! \langle #2 \vert}

\newcommand{\Tr}[1]{\textrm{Tr}\,\! #1}

\begin{document}
\title{Optimal quantum estimation of the coupling constant of Jaynes-Cummings interaction}
\author{Marco G. Genoni\inst{1}\fnmsep\thanks{\email{m.genoni@imperial.ac.uk}} \and Carmen Invernizzi\inst{2} }
\institute{QOLS, Blackett Laboratory, Imperial College London, 
London SW7 2BW, UK \and Dipartimento di Fisica, Universit\`a degli Studi di Milano,
I-20133 Milano, Italia}
\abstract{
We address the estimation of the coupling constant of the Jaynes-Cummings Hamiltonian 
for a coupled qubit-oscillator system. 
We evaluate the quantum Fisher Information (QFI) for the 
system undergone the Jaynes-Cummings evolution, considering
that the probe initial state is prepared in a Fock state for the oscillator 
and in a generic pure state for the qubit; we obtain that the
QFI is exactly equal to the
number of excitations present in the probe state.
We then focus on the two subsystems, namely the qubit and the oscillator alone,
deriving the two QFIs of the two reduced states, and comparing them with 
the previous result. 
Next we focus on feasible measurements on the system, and we
 find out that if  population measurement on the qubit 
and Fock number measurement on the oscillator are performed together, the 
Cramer-Rao bound is saturated, that is the corresponding Fisher Information (FI)
is always equal to the QFI.
We compare also the performances of these measurements 
performed alone, that is when one of the two subsystem is ignored.
We show that, when the qubit is prepared in either the ground or the excited state, 
the local measurements are still optimal. }
\maketitle
\section{Introduction}
\label{intro}

The Jaynes-Cummings (JC) model \cite{JCmodel} is one of the paradigmatic examples
of {\em hybrid systems}, where a two-level system, modelled by a spin-$1/2$, is coupled to a
 quantized mode of a harmonic oscillator. This is the typical situation in quantum 
 electrodynamics (QED) \cite{Haroche} where a single atom can be coupled to a cavity mode, 
 for example in the microwave \cite{microwave,microwave2} and in the optical 
 \cite{optical} regime. 
The same model describes accurately other interesting physical systems, such as a single ion 
or a neutral atom in a trap \cite{Leibfried}, and the interaction of artificial atoms with resonators in circuit QED systems \cite{circuit00,circuit01}. All these {\em hybrid systems}
are considered essential for the development of quantum information processing 
and in general of future quantum technologies \cite{wiring,qinternet}. 
The aim of quantum information is to characterize the peculiar properties
of quantum systems and exploit them to perform tasks that would be not achievable in
a classical context \cite{NC}.
For these purposes it is necessary to characterize the value of quantities that
are not directly accessible either in principle or due to experimental impediments.
This is the case of relevant quantities like phase \cite{metrology,qphase1,qphase2}, 
entanglement  \cite{estent,estentEXP}  or temperature \cite{temperature}, 
that cannot correspond to any quantum observable
or also the coupling constant of different kinds of interactions \cite{JIsing,loss,Kerr}.
In this paper we apply the techniques of local quantum estimation theory (QET) \cite{Helstrom,Braunstein,Brody,lqe} to the problem of estimating the coupling constant 
of the JC Hamiltonian.
In this framework, the characterization of the estimation of the parameter is provided by the Fisher information (FI) which represents an infinitesimal distance between probability distributions and gives the ultimate precision attainable by an estimator via the Cramer-Rao theorem. Its quantum counterpart, the quantum Fisher information (QFI), is related to the degree of distinguishability of a quantum state from its neighbours \cite{Bures,Uhlmann,Wootters} and gives the ultimate bound to the precision on the estimate allowed by quantum mechanics. \\
In our estimation problem we consider the initial probe state prepared in a Fock state,
as regards the oscillator and in a generic pure state, as regards the qubit. We address
the overall estimation properties, by evaluating the QFI for the whole system undergone 
the JC evolution. We also focus on the two subsystems alone, i.e. we consider the 
qubit subsystem obtained as the partial trace over the harmonic oscillator and 
{\em viceversa}, and we evaluate the QFI of the corresponding reduced states, in order to 
identify how much information about the parameter of interest is contained in each subsystem.
Moreover we consider as possible feasible measurements on the coupled system, 
the measurement of the population of the  excited state performed on 
the qubit and a measurement of the Fock number performed on the harmonic oscillator. 
We evaluate the FI for the collective measurement and observe that it allows one to achieve 
the ultimate bound on the precision  given by the 
quantum Cramer-Rao bound. Finally we consider the FI for the same measurements performed 
on the two corresponding subsystems alone, that is when one of the two subsystems
is ignored. We show that, when the qubit is prepared in either the ground or the excited state, 
both single local measurements are still optimal, that is the corresponding FIs are 
equal to the QFI of the whole systems. \\
The paper is structured as follows: in Sec. \ref{sec:1} we describe the JC model and 
the unitary dynamics for the coupled system. In Sec. \ref{sec:2} we review some 
concepts of QET and illustrate the quantum Cramer-Rao bound focusing to the case 
of a pure-state unitary family. In Sec. \ref{sec:3} we show in details our results  and 
finally, Sec. \ref{conclusions} gives some concluding remarks.  
\section{The model}\label{sec:1}
The JC model describes the interaction between a single-mode bosonic field
and a two-level system (qubit). Its dynamics is exactly solvable and the model has been widely investigated experimentally \cite{Haroche}.

The bosonic field is described upon introducing
an annihilation and a creation operator, satisfying $[a,a^\dagger] = \mathbbm{1}$,
and the corresponding Fock states $\{ |n\rangle \}_{n=0}^\infty\:$, {\em i.e.} the eigenstates
of the number operator $N=a^\dag a$, which provide a basis
of the infinite-dimensional  Hilbert space. In the rest of the manuscript we will adopt 
the quantum optics terminology calling {\em photons} the bosonic excitations,
however the results presented are also valid for the other physical 
settings described by the JC model. \\
The two-level system (qubit) is characterized by a 
ground state $|g\rangle$ and an excited state $|e\rangle$ which are eigenstates
of the Pauli operator $\sigma_z$ and in turn form 
a basis for the two-dimensional Hilbert space of the qubit. Any pure
state of the whole system composed by the bosonic field and the qubit can be written as 
$|\Psi\rangle = \sum_{j=g,e} \sum_n
c_{j,n} |j,n\rangle$, where we denote with $|j,n\rangle=|j\rangle \otimes |n\rangle$
the tensor product between the state of the qubit and the state of the field. 

The JC Hamiltonian reads
\begin{align}
\mathcal H = \frac{\hbar \omega_q}{2} \sigma_z + \hbar\omega_f \: a^\dag a + \frac{\hbar g}{2} 
(\sigma_+ a + \sigma_- a^\dagger ) 
\end{align}
where the operators $\sigma_\pm$ are the qubit ladder operators 
$\sigma_+ =\ketbra{e}{g}$ and $\sigma_-=\ketbra{g}{e}$. The 
two operators $\sigma_+ a$ and $\sigma_- a^\dagger$ correspond 
respectively to a transition from the lower level $|g\rangle$ to the 
upper level $|e\rangle$ together with the emission of a photon, 
and the transition from $|g\rangle$ to $|e\rangle$ together with the 
annihilation of a photon. It is clear that the interaction couples, for a given integer 
$n$, the states $\ket{e,n}$ and $\ket{g,n+1}$. 

Upon choosing a suitable rotating frame and by considering the resonance 
condition $\omega_q = \omega_f$, one rewrites the Hamiltonian $\mathcal H$
in the so called interaction picture as
\begin{align}
\mathcal H_{JC}= \frac{\hbar g}{2} (\sigma_+ a+ \sigma_-a^\dagger ) .
\end{align}
The corresponding evolution unitary operator for an interaction time $\tau$ reads
\begin{align}
U_{JC}(\Omega) = \exp \left( -\frac{i}{\hbar}\mathcal H_{JC} \tau \right) = \exp (- i \Omega G )
\end{align}
where we defined  
\begin{align}
G_{JC}=\frac{\sigma_+a + \sigma_-a^\dagger }{2} \:\:\:\: \textrm{and} \:\:\:\: \Omega = g \tau.
\end{align}
The parameter $\Omega$ is the quantity of interest when we want
to control the JC dynamics and in the following sections 
we will focus on its estimation properties.

In our treatment we assume that at time $t=0$ the probe state is prepared in 
a pure state and no initial correlations between the 
qubit and the field are present, in formula
\begin{align}
\varrho(0)=|\Psi_0\rangle\langle\Psi_0|
\end{align}
with $|\Psi_0\rangle = |\psi_q\rangle \otimes |\psi_f\rangle$. In particular
the qubit at time $t=0$ is prepared in a pure superposition of 
ground and excited states,
\begin{align}
|\psi_q\rangle = \cos\frac{\theta}{2} |e\rangle+ \sin\frac{\theta}{2} |g\rangle \label{eq:superpos}
\end{align}
while the bosonic field is prepared in a Fock state $|\psi_f\rangle=|n\rangle$. 
Notice that preparation of Fock states have been proposed theoretically and proved experimentally
both in cavity QED, ion trapping and circuit-QED systems \cite{Haroche,Cirac,Wineland,Walther,Bertet,HaroceFB,circuit,circuit2}. 
Given this probe preparation, the evolution of the system reads
\begin{align}
\varrho(\Omega)=U_{JC}(\Omega)\varrho(0)U_{JC}^\dagger(\Omega).
\end{align}
and upon tracing the evolved state over the bosonic field or the qubit degrees of freedom,
we obtain the states
\begin{align}
\varrho_q(\Omega)=\Tr_f[U_{JC}(\Omega)\varrho(0)U_{JC}^\dagger(\Omega)] \\
\varrho_f(\Omega) =\Tr_q[U_{JC}(\Omega)\varrho(0)U_{JC}^\dagger(\Omega)] 
\label{eq:subsystems}
\end{align}
which describe respectively the qubit state and the harmonic
oscillator subsystem state at time $t$. 

In details, the reduced qubit
density operator reads in the basis $\{ |e\rangle, |g\rangle\}$ as 
\begin{align}
\varrho_q(\Omega)=\left (\begin{array}{cc}
\varrho_{ee} &\varrho_{eg} \\
\varrho_{ge} &\varrho_{gg}
 \end{array}\right ), \label{eq:qubitreduced}
\end{align}
where
\begin{align}
\varrho_{gg}=&\cos^2\frac \theta 2\sin^2 \left(\frac{\Omega}{2} \sqrt{n+1}\right) +\sin^2\frac\theta 2 \cos^2\left (\frac{\Omega}{2}\sqrt n\right)  \nonumber \\
\varrho_{ee} =& 1-\varrho_{gg}   \nonumber \\
\varrho_{eg}=& \frac 1 2\sin\theta\cos\left(\frac \Omega 2\sqrt n\right)\cos\left(\frac\Omega 2\sqrt{n+1}\right)
=\varrho_{ge}  \label{eq:qubitreduced3}.
\end{align}
On the other hand, the reduced density operator for the bosonic field is diagonal in the Fock
basis:
\begin{align}
\varrho_f(\Omega) = p_{n-1} |n-1\rangle\langle n-1| + p_n |n\rangle\langle n| + p_{n+1} 
|n+1\rangle\langle n+1|,\label{eq:bosonreduced}
\end{align}
where
\begin{align}
p_{n-1} &= \sin^2 \frac\theta 2\left( \frac{1 - \cos \left(\Omega \sqrt{n} \right)}{2}\right) \nonumber \\
p_n &= \frac12 \left( 1 + \cos^2 \frac\theta 2 \cos\left (\Omega \sqrt{n+1}\right) + 
\sin^2 \frac\theta 2 \cos\left (\Omega \sqrt{n}\right)  \right) \nonumber \\
p_{n+1} &= \cos^2 \frac\theta 2 \left(\frac{1 - \cos \left(\Omega \sqrt{n+1} \right)}{2}\right).
 \label{eq:bosonreduced2}
\end{align}
\section{Local quantum estimation theory}\label{sec:2}
In this section we review the basic concepts of local quantum estimation theory that 
will be applied later to the qubit-oscillator system.

An estimation problem consists into choosing a measurement 
somehow related with the parameter of interest and then defining 
an estimator, i.e. a function from the set of the measurement outcomes 
to the parameter space, in order to infer the value of the quantity that we 
want to estimate. Classically, given the conditional probability
$p(j|\Omega)$ of measuring the outcome $j$ when the value to be 
estimated is $\Omega$, optimal estimators are those saturating the 
Cramer-Rao inequality, which establishes that the variance $\hbox{Var}(\Omega)$ 
of any unbiased estimator is lower bounded by 
\begin{align}
\hbox{Var}(\Omega)\geq \frac{1}{M F(\Omega)} ,
\end{align}
where $M$ is the number of measurements of the sample and $F(\Omega)$ 
the Fisher information (FI)
\begin{align}
F(\Omega)=\sum_{j} p(j|\Omega)[\partial_\Omega \ln p(j|\Omega)]^2. \label{eq:FI}
\end{align}
In quantum mechanics, according to the Born rule one has 
$p(j|\Omega)=\Tr[\varrho_\Omega \Pi_j ]$, where $\varrho_\Omega$ is a 
family of quantum states which depend on the parameter $\Omega$ 
and the operators $\{\Pi_j\}$ are the elements of the probability operator-valued
measure (POVM) describing the quantum measurement. By defining the symmetric
logarithmic derivative (SLD) operator $L_\Omega$ by means of the following equation
\begin{align}
\frac{\partial\varrho_\Omega}{\partial\Omega} = \frac{ L_{\Omega} \varrho_\Omega + \varrho_\Omega L_\Omega}2,
\end{align}
the classical Fisher Information in Eq. (\ref{eq:FI}) can be rewritten as 
\begin{align}
F(\Omega) = \sum_j \frac{{\rm Re}(\Tr[\varrho_\Omega \Pi_j L_\Omega ])^2}{\Tr[\varrho_\Omega \Pi_j]}
\end{align}
which establishes the maximum precision for the estimation of the parameter $\Omega$ 
for a fixed measurement $\{\Pi_j\}$. Moreover, maximizing the FI over
all the possible measurements, one can show that
\begin{align}
F(\Omega) \leq H(\Omega) = \Tr [\varrho_\Omega L_\Omega^2 ] .
\end{align}
The quantity $H(\Omega)$ is called quantum Fisher information (QFI) and
define the corresponding quantum Cramer-Rao bound
\begin{align}
\hbox{Var}(\Omega)\geq\frac{1}{M H(\Omega)},
\end{align}
which gives the ultimate limit to the precision allowed from quantum mechanics for the estimation of 
the parameter $\Omega$ labelling a 
given quantum statistical model $\varrho_\Omega$.

When the set of quantum states $\varrho_\Omega$ is given in a diagonal form $\varrho_\Omega= \sum_k \lambda_k |\phi_k\rangle \langle \phi_k |$, the QFI can be evaluated as 
\begin{align}
H(\Omega) = \sum_k \frac{(\partial_\Omega \lambda_k)^2}{\lambda_k} + 
2 \sum_{k,l} \frac{(\lambda_k - \lambda_l)^2}{\lambda_k + \lambda_l} 
|\langle\phi_l | \partial_\Omega \phi_k\rangle|^2.
 \label{eq:QFIr}
\end{align}
A particular case is given when the family of $\varrho_\Omega$ is a unitary pure-state family, 
\begin{align}
\varrho_{\Omega} = U_\Omega |\Psi_0\rangle\langle \Psi_0|U_\Omega^\dagger \qquad 
U_\Omega = \exp ( - i G \Omega) 
\end{align}
where $G$ is the generator of the transformation and $|\Psi_0\rangle$
the initial pure probe state. In this case one can prove that the QFI is independent 
on the actual value of the parameter $\Omega$ and it is proportional to the fluctuations
of the generator on the probe state, i.e.
\begin{align}
H(\Omega) &= 4\: \langle \Delta^2 G \rangle_0 \nonumber \\
&= 4\left( \langle \Psi_0 | G^2 |\Psi_0\rangle - \langle \Psi_0 | G |\Psi_0\rangle ^2 \right)
\label{eq:unitaryfam}
\end{align}
One can thus rewrite the quantum Cramer-Rao bound and read it
as a Heisenberg-like uncertainty relation
\begin{align}
\hbox{Var}(\Omega )\: \langle \Delta^2 G \rangle_0 \geq \frac{1}{4 M}.
\end{align}
\section{Results}\label{sec:3}
In this section we report the results for the estimation of the parameter $\Omega$ in the 
JC model. 
We first derive the ultimate limit to the estimation
by evaluating the QFI for the whole coupled system composed by the qubit and the harmonic oscillator. 
We then consider the two subsystems alone, by performing a partial trace
either on the bosonic field or the qubit, and evaluate the QFI of the reduced 
states $\varrho_q(\Omega)$ and $\varrho_f(\Omega)$ given respectively in Eqs. 
(\ref{eq:qubitreduced}) and (\ref{eq:bosonreduced}) .
We also consider the performances of two feasible measurements: 
the photon number measurement on the bosonic field and the population measurement 
on the qubit. 
We first evaluate the FI for the collective measurement on the total system and
finally, we consider the FI for the same measurements performed on the two subsystems 
alone, namely the FI for population measurements on the qubit subsystem and the FI for 
photon number measurements on the oscillator subsystem alone. 
\subsection{Quantum Fisher Information}
In the following we evaluate the QFI for the pure state $\varrho(\Omega)$ obtained upon choosing
as a probe state $|\Psi_0\rangle = |\psi_q\rangle \otimes |n\rangle$, where the state of the qubit is a general superposition $|\psi_q\rangle$ as in Eq. (\ref{eq:superpos}) and 
the field is prepared into a Fock state $|n\rangle$. \\
We consider the evolution of the whole system and
assume that both the degrees of freedom of the qubit and of the field are accessible and
measurable, i.e. our statistical model is a pure state unitary family 
and the corresponding QFI for the parameter $\Omega$ can be evaluated as in Eq. (\ref{eq:unitaryfam})
with the generator $G_{JC}=(a \sigma_+ + a^\dagger \sigma_-)/2$. 
The calculation leads to
\begin{align}
H(\Omega) = n+ \cos^2\frac\theta{2}.
\end{align}
Note that the QFI $H(\Omega)$ has the minimum value for $\theta=\pi$, where $H(\Omega)=n$,
and the maximum is $H(\Omega)=n+1$ for $\theta=0$. Thus the optimal preparation for the qubit
state which gives the maximum value of the QFI, corresponds to the excited state $|\psi_q\rangle=|e\rangle$. In particular one can observe that the QFI $H(\Omega)$ of the total system is equal to 
the total number of excitations of the probe state, 
\begin{align}
H(\Omega) = \langle \Psi_0 | E |\Psi_0\rangle \:\:\: {\rm where} \:\:\:\:
E=a^{\dag}a + \sigma_+\sigma_-,
\end{align}
showing that the more excitations we have, the more precise will
be the estimation of the JC coupling constant.\\
We now consider the case where the degrees of freedom of one of the
two subsystems are not accessible. The reduced states of the qubit and the bosonic field 
are given respectively in
Eqs. (\ref{eq:qubitreduced}) and (\ref{eq:bosonreduced}) and the corresponding QFIs
are denoted as $H_q(\Omega)$ and $H_f(\Omega)$. 
We evaluate them by means 
of Eq. (\ref{eq:QFIr})
\begin{figure}
\includegraphics{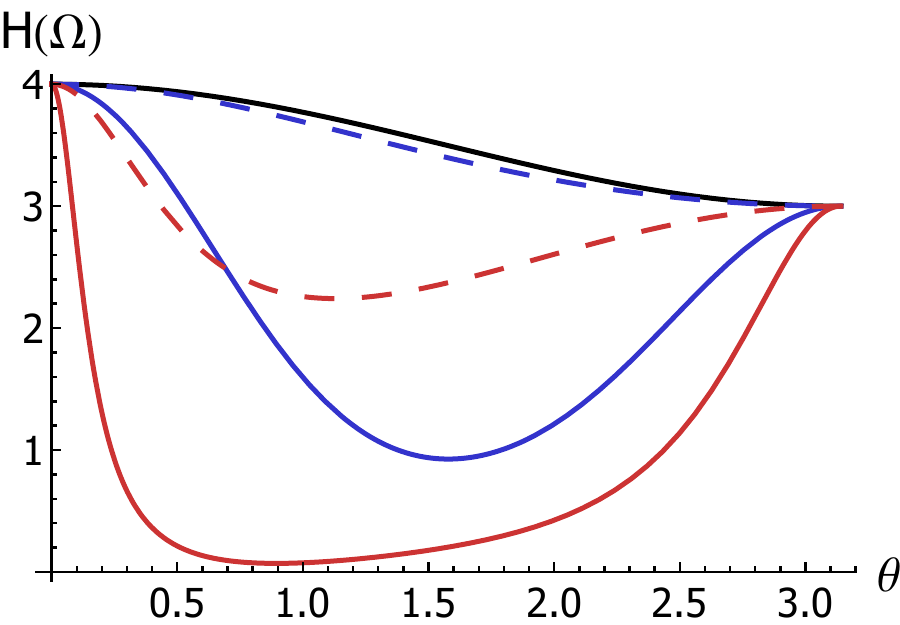}
\caption{
QFIs of the two subsystems as a function of $\theta$ with $n=3$ and for different
values of the parameter $\Omega$ (blue lines, $\Omega=1.0$; red lines, $\Omega=1.5$). The 
dashed lines denote the QFI $H_f(\Omega)$ of the bosonic field, while the solid 
lines denote the QFI $H_q(\Omega)$ of the qubit subsystem. The black solid line on top,
represents the QFI $H(\Omega)$ of the qubit-oscillator total system. 
\label{f:QFI}
}
\end{figure} 
and plot their behaviour in Fig. \ref{f:QFI} as a function of $\theta$ and for different
values of the parameter $\Omega$. We observe that the bosonic field alone
contains more information about the parameter $\Omega$ than the qubit subsystem,
that is
\begin{align}
 H_f(\Omega) \geq H_q(\Omega).
\end{align}
Moreover, we also observe that if the qubit is prepared in either the 
ground or excited state, the values of the QFI of the total system $H(\Omega)$
coincides to the QFIs of the two reduced subsystems.
In formula we have that, for all the possible values
of $\Omega$,
\begin{align}
H_q(\Omega)[\theta=0]=H_f(\Omega)[\theta=0] &=n+1 \\
H_q(\Omega)[\theta=\pi]=H_f(\Omega)[\theta=\pi] &=n. 
\end{align} 
This result shows that, for particular choices of the input state, a measure 
performed on one of the subsystems, either the qubit or the oscillator,
can attain the ultimate precision on the estimate of the parameter $\Omega$ given by the QFI $H(\Omega)$.
\subsection{Fisher information for Fock and population measurements}
In order to address the performances of feasible measurements for the 
estimation of the parameter $\Omega$, 
we consider the FI for the photon number measurement 
on the bosonic field and the population measurement on the qubit system. 
If we consider a collective measurement performed on both subsystems, the 
non-zero conditional probabilities that we have to take into account are
\begin{align}
p(e, n+1 | \Omega) &= \cos^2 \frac\theta 2 \left( \frac{1-\cos\left(\Omega \sqrt{n+1} \right) }{2}\right) \\
p(e, n | \Omega) &= \sin^2 \frac\theta 2 \left( \frac{1+\cos\left(\Omega \sqrt{n} \right) }{2}\right) \\
p(g, n | \Omega) &= \cos^2 \frac\theta 2 \left( \frac{1+\cos\left(\Omega \sqrt{n+1} \right) }{2}\right) \\
p(g, n-1 | \Omega) &= \cos^2 \frac\theta 2 \left( \frac{1-\cos\left(\Omega \sqrt{n} \right) }{2}\right),
\end{align}
where $p(j,n|\Omega)$ denotes the conditional probability of obtaining the qubit
in the state $j$ and the bosonic field with $n$ excitations, 
when the parameter has the value $\Omega$.
The corresponding Fisher information is evaluated by means of Eq. (\ref{eq:FI}) giving
\begin{align}
F(\Omega) = H(\Omega) = n + \cos^2 \frac\theta 2.
\end{align}
Therefore the Fisher information $F(\Omega)$ for 
the Fock and population measurements saturates
the quantum Cramer-Rao bound i.e. the measurement considered  is always optimal.

We now study the two measurements separately, i.e. we evaluate  the Fisher information
for population measurements on the qubit $F_q(\Omega)$ and the Fisher 
information for the photon number measurements on the harmonic oscillator $F_f(\Omega)$. 
In particular we want to compare them with the corresponding QFIs for the two subsystems $H_q(\Omega)$ and $H_f(\Omega)$. 
The population measurement
on the qubit corresponds to measure the Pauli operator $\sigma_z$, 
whose eigenstates are indeed $\{ |e\rangle, |g\rangle \}$. 
The conditional probabilities for the qubit to be found in the ground or the 
excited state are given by the diagonal elements of the reduced state
$\varrho_q(\Omega)$, that is
\begin{align}
p(g |\Omega)&= \varrho_{gg} = \cos^2\frac \theta 2\sin^2 \left(\frac{\Omega}{2} \sqrt{n+1}\right) +\sin^2\frac\theta 2 \cos^2\left (\frac{\Omega}{2}\sqrt n\right)  \\
p(e|\Omega) &= \varrho_{ee} =1-\varrho_{gg}.
\end{align}
On the other hand, the non-zero conditional probabilites for the number measurement
on the bosonic fields are 
\begin{align}
p(n-1|\Omega) = p_{n-1} &= \sin^2 \frac\theta 2\left( \frac{1 - \cos \left(\Omega \sqrt{n} \right)}{2}\right) \nonumber \\
p(n|\Omega) = p_n &= \frac12 \left( 1 + \cos^2 \frac\theta 2 \cos\left (\Omega \sqrt{n+1}\right) + 
\sin^2 \frac\theta 2 \cos\left (\Omega \sqrt{n}\right)  \right) \nonumber \\
p(n+1|\Omega) = p_{n+1} &= \cos^2 \frac\theta 2 \left(\frac{1 - \cos \left(\Omega \sqrt{n+1} \right)}{2}\right)
\end{align}
\begin{figure}
\includegraphics{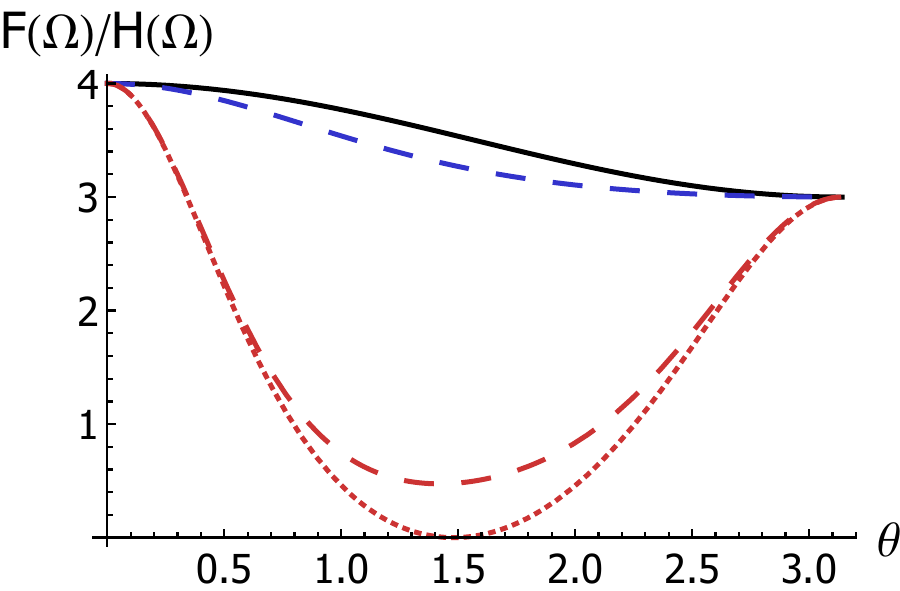}
\caption{
(From top to bottom) Black solid line: QFI $H(\Omega)$ and FI $F(\Omega)$
for collective Fock and population measurement. Dashed blue line: QFI $H_f(\Omega)$ and FI $F_f(\Omega)$ for Fock measurement on the harmonic oscillator subsystem alone. Dashed red line: QFI $H_q(\Omega)$. Dotted red line: FI $F_q(\Omega)$ for population measurement on the qubit subsystem alone. 
The parameters considered are $\Omega=1.25$ and $n=3$.
\label{f:FI}
}
\end{figure} 
The corresponding FIs, $F_q(\Omega)$ and $F_f(\Omega)$, 
are evaluated by means of Eq. (\ref{eq:FI}) and are plotted
in Fig. \ref{f:FI}. 
Since the field subsystem state $\varrho_f(\Omega)$ is diagonal in the 
Fock number basis, we have that the Fock number measurement is always 
optimal, i.e. the Fisher information is equal to the quantum Fisher information
of the state $\varrho_f(\Omega)$
\begin{align}
F_f(\Omega) = H_f(\Omega) \qquad \forall \: \: \Omega,\theta, n. 
\end{align}
On the other hand, the population measurement
is in general not optimal but it saturates the subsystem quantum Cramer-Rao bound 
when $\theta=0$ or $\theta=\pi$, that is when the qubit is prepared 
respectively in the excited state $|e\rangle$ or
in the ground state $|g\rangle$. In these special cases
all the quantities we have evaluated so far
are equal, that is the FIs for the measurements on the two subsystems are 
equal not only to the corresponding QFIs of the reduced states $H_f(\Omega)$ 
and $H_q(\Omega)$, but also to the QFI of the total system $H(\Omega)$. 
In formula we have that
\begin{align}
F_q(\Omega) = F_f(\Omega) = H(\Omega) = 
\begin{cases}
n+1 \:\:\: \textrm{if}\:\: \theta=0 \\
n \qquad\:\: \textrm{if}\:\: \theta=\pi
\end{cases}
\end{align}
This interesting feature suggests that for these particular
choices of the qubit preparation, in order to attain 
the ultimate limit posed by the quantum Cramer-Rao bound, we can decide 
to measure only one of the two subsystems neglecting the other
degrees of freedom. 
\section{Conclusions}\label{conclusions}
In this paper we addressed the estimation of the coupling constant
of the Jaynes-Cummings Hamiltonian considering
as an input probe state the field prepared into a Fock state $|n\rangle$ and
the qubit prepared in a generic superposition of excited and ground state.\\
We derived the QFI for the whole system, i.e. the qubit-oscillator system evolved under 
the JC Hamiltonian and observed that it is equal
to the number of excitations of the input probe state and independent on the 
value of the parameter to be estimated. We then derived the QFI
for the two subsystems, i.e. the QFI of the qubit and the oscillator state. 
In this case we have found that in principle the parameter is better estimated 
when the qubit subsystem is ignored, rather than the bosonic field.\\
We then considered a feasible detection scheme where 
a measurement of the population of the excited state is performed on 
the qubit and a measurement of the Fock number is performed on the harmonic oscillator. 
The FI for the corresponding collective measurement
turned out to be equal to the corresponding QFI, saturating the 
quantum Cramer-Rao bound and providing the optimal estimate of the parameter.\\
Finally we considered the FI for the same measurements performed separately on the two subsystems alone, namely by ignoring the degrees of freedom of one subsystem. 
The surprising result is that, if the qubit is prepared either in the ground or in the
excited state, both the measurements performed on the single subsystems,
and ignoring the remaining one, are optimal. This result is relevant in many practical 
situations where one of the two subsystems is not experimentally accessible.
Moreover since the bound obtained does not depend on the value of the parameter, 
it is not necessary to tune the measurement by means of two-step or adaptive 
estimation strategies in order to attain the optimality.\\
As we stressed above, our study shows the existence of a link between the estimation 
of the JC coupling constant and the amount of excitations present in the 
probe state. Since the preparation of a high-number Fock state is still experimentally
challenging, as a future outlook, it would be interesting
to study different preparations of the probe states, for example
by considering the bosonic field in a coherent state with a high number
of photons. It will be relevant in this case, also to understand if the measurements
that in this case has been proved to be optimal, remain optimal with such a
 different preparation of the probe.
\section{Acknowledgments}
The authors acknowledge useful discussions with Stefano Olivares and Matteo Paris.
MGG acknowledges support from UK EPSRC (grant EP/I026436/1).

\end{document}